\documentclass[prd,aps,12pt]{revtex4}
\usepackage{latexsym}
\usepackage{amssymb}
\usepackage[dvips]{graphicx}
\usepackage{color}
\usepackage{graphicx}
\begin{document}
\newcommand{\beq}{\begin{equation}}
\newcommand{\eeq}{\end{equation}}
\newcommand{\ie}{{\sl i.e\/}}
\newcommand{\half}{\frac 1 2}
\newcommand{\lag}{\cal L}
\newcommand{\ove}{\overline}
\newcommand{\et}{{\em et al}}
\newcommand{\Prd}{Phys. Rev D}
\newcommand{\Prl}{Phys. Rev. Lett.}
\newcommand{\Plb}{Phys. Lett. B}
\newcommand{\Cqg}{Class. Quantum Grav.}
\newcommand{\Grg}{Grav....}
\newcommand{\Np}{Nuc. Phys.}
\newcommand{\Fp}{Found. Phys.}
\renewcommand{\baselinestretch}{1.2}

\title{A possible interaction between uncharged particles and the Electromagnetic field}
\author{M. Novello and C.E.L. Ducap}
\affiliation{Centro de Estudos Avan\c{c}ados de Cosmologia (CEAC/CBPF) \\
 Rua Dr. Xavier Sigaud, 150, CEP 22290-180, Rio de Janeiro, Brazil}

\date{\today}

\begin{abstract}
Maxwell Electrodynamics can be described either in Minkowski space-time or in a dynamically equivalent way in a curved geometry constructed in terms of the electromagnetic field. For this the field must have a superior bound limited by a maximum value of its energy.  After showing such geometric equivalence we investigate the possibility that new processes dependent on the metric structure may appear. In particular for very high values of the field a direct coupling of uncharged particles to the electromagnetic field is predicted.
\end{abstract}

\vskip2pc
 \maketitle

\section{Introduction}

We would like to point out from the very beginning that the geometric phenomenon in the realm of Electrodynamics that we describe in this work has only few similarities with the crucial ideas of geometry that are present in general relativity \cite{einsteingr}  or in other analogous geometric approaches like, for instance, the one produced by Gordon \cite{gordon} \cite{novellovisservolovik} or in the recent theory of metric relativity \cite{novelloeduardo}. In those theories the methodology can be synthesized in a single idea: to eliminate a force field that produces acceleration on bodies. This is accomplished using a  geometric formulation that describes accelerated paths in Minkowski background as geodesics in another geometry. Nothing similar here. Thus we are not on the way to produce a geometric formulation of the Electromagnetic field in the Einstein\rq s sense,  but instead we use a specific form of geometry, constructed with the electromagnetic field, to show an equivalent way to describe its dynamics. In other words, Maxwell electrodynamics described in a Minkowski space-time can be equivalently described as a specific non-linear theory in an associated electromagnetic geometry (e-geometry for short). We analyze the behavior of test particles in this formulation. We are then led to examine a further consequence of such geometric way assuming that this effective metric is universal, that is it pervades the space-time for all kinds of particles. Then an immediate question arises: how to conciliate this proposal with the observation that there exist particles which are inert  in an electromagnetic field? The solution of this is related to the intensity of the field. Indeed, the effects of such geometric formulation are important only for very high values of the field, in the verge of its maximum possible value, which we name $ \beta.$ We analyze the total transparency of uncharged particles when passing through a domain of space-time filled with an electromagnetic field and question the absolute restriction that such property imposes.

In the first moment, we will show that the presence of an equivalent metric associated to the space-time can be used to describe the electromagnetic field. Having defined this electromagnetic metric (e-metric, for short) two ways to couple fields to the electromagnetic field appear:
\begin{itemize}
\item{The standard minimal coupling principle which changes the derivative $ \partial_{\mu} $ into the form $ \partial_{\mu} - i \, e \, A_{\mu}; $}
\item{Minimal coupling with the e-metric which changes $ \partial_{\mu} $ into a covariant derivative $ \nabla_{\mu} $ defined in terms of the connection of the metric.}
\end{itemize}
A charged particle interacts through both channels, an uncharged particle only trough the e-metric channel.

The e-metric contains not only the electromagnetic tensor $ F_{\mu\nu} $ but also a constant $ \beta.$ We shall see that this constant is interpreted as the maximum possible value for the electromagnetic field. All new results in this work rely upon the existence of an upper limit. If this value is taken to be infinite, then the e-metric becomes identical to the Minkowski background and we recover that uncharged bodies do not interact with the electromagnetic field. In this vein, the new properties concerning the possibility of a new channel of interaction with the electromagnetic field which does not conflict with standard observations  depends upon the smallness of the quantity $ 1/ \beta$ that controls the intensity of the e-metric.

\section{The correspondence bridge}

  Recently it has been argued \cite{novelloeduardo}  that the standard Maxwell Electrodynamics in a Minkowski background can alternatively be described as a nonlinear theory described in a curved space endowed with a metric constructed with the electromagnetic field itself. These two dynamics are just one and the same. We present a rather simple but direct proof of this statement.

Let $ S_{M}$ represent the free part of Maxwell\rq s action in Minkowski space-time:
$$ S_{M} = - \, \int \, \sqrt{- \eta} \, F \, d^{4}x $$
where $  \eta$ is the determinant of the Minkowski metric,
$$ F = F_{\mu\nu} \, F^{\mu\nu} =  F_{\alpha\mu} \, F_{\beta\nu} \, \eta^{\alpha\beta} \, \eta^{\mu\nu} $$
and for completeness we name the other invariant $ G $ by
$$ G = F^{*}_{\mu\nu} \, F^{\mu\nu} = \frac{1}{2} \,\eta^{\alpha\beta\mu\nu} \, F_{\alpha\beta} \, F_{\mu\nu} $$
where $ \eta^{\alpha\beta\mu\nu} = - \varepsilon^{\alpha\beta\mu\nu} / \sqrt{-\eta}.$

Define the electromagnetic metric as \cite{others}
\begin{equation}
 \hat{e}^{\mu\nu} =  \eta^{\mu\nu} - \, \frac{1}{\beta^{2}} \, \Phi^{\mu\nu}
 \label{7set1}
 \end{equation}
in which the constant $ \beta^{2}$ has the dimension of $ [F] $ and $\Phi^{\mu\nu} = F^{\mu\alpha} \, F_{\alpha}{}^{\nu}.$

The determinant of the e-metric is given by

$$ \sqrt{- \hat{e}} = \frac{\sqrt{- \eta}}{U}$$
where

$$ U= \, 1 + \frac{F}{2 \, \beta^{2}} - \frac{G^{2}}{16 \, \beta^{4}}.$$
In the case of a pure electrostatic field $ U = 1 - E^{2} / \beta^{2}.$ From the form of the determinant and to keep the geometry (eq. \ref{7set1}) well-behaved the quantity $ U $ should not change sign. This yields a limitation for the field, that is
$$ E^{2} < \beta^{2}.$$
We can thus attribute to the constant $ \beta$ the meaning of the maximum possible value for the field \cite{bi}.  For latter application, we explicit the identity

$$ \Phi_{\mu}{}^{\alpha} \, \Phi_{\alpha\nu} = \frac{1}{16} \, G^{2} \, \eta_{\mu\nu} - \frac{1}{2} \, F \, \Phi_{\mu\nu}.$$

We assume that the e-metric is riemannian, that is, there is a covariant derivative constructed with this metric in such way that it obeys the metricity condition
$$ \hat{e}^{\mu\nu}{}_{; \,\alpha} = 0$$ which allows to define a connection

$$ \hat{\Gamma}^{\alpha}_{\mu\nu} = \frac{1}{2} \, \hat{e}^{\alpha\lambda} \, \left( \hat{e}_{\lambda\mu , \nu} + \hat{e}_{\lambda\nu, \mu} - \hat{e}_{\mu\nu ,\lambda}\right)$$
where  a comma means ordinary derivative. We use a hat to characterize objects that are represented in the e-metric. The inverse metric defined as
$$  \hat{e}^{\alpha\nu} \, \hat{e}_{\nu\beta} = \delta^{\alpha}_{\beta} $$
is given by
$$ \hat{e}_{\mu\nu} = A \, \eta_{\mu\nu} + B \, \Phi_{\mu\nu} $$
where

\begin{equation}
 A = \frac{2 + F/\beta^{2}}{2 \, U}
 \label{4out3}
 \end{equation}
 \begin{equation}
  B =  \frac{1}{\beta^{2} \, U}.
  \label{4out5}
  \end{equation}

We remind that for any non-hat tensorial quantity all properties concerning the relationship between covariant and contra-variant indices are performed by Minkowski metric $ \eta_{\mu\nu}.$ The effect of the map from Minkowski geometry into the e-metric does not change the covariant form of the electromagnetic tensor, that is
 $$ \hat{F}_{\mu\nu} = F_{\mu\nu}.$$
 It then follows for the contra-variant form

\begin{equation}
 \hat{F}^{\mu\nu} = \hat{e}^{\mu\alpha} \, \hat{e}^{\nu\beta} \, F_{\alpha\beta} = \left( \Sigma \, F^{\mu\nu} + \Pi \, F^{*}{}^{\mu\nu} \right)
 \label{6set1}
 \end{equation}

Consequently the associated invariants $ \hat{F} $ and $ \hat{G}$ are related to their corresponding ones in the Minkowski space by

\begin{equation}
\hat{F} =  \, \left( \Sigma \, F + \Pi \, G\right)
\label{1}
\end{equation}
and
\begin{equation}
\hat{G} = \, U \, G
\label{2}
\end{equation}
where
\begin{equation}
 \Sigma = \, 1 + \frac{F}{\beta^{2}} + \frac{1}{4 \, \beta^{4}} ( F^{2} + \frac{G^{2}}{4})
 \label{11set1}
 \end{equation}
and
 \begin{equation}
 \Pi =  \frac{1}{2 \, \beta^{2}} \, G \, ( \frac{1}{4 \, \beta^{2}} \, F + 1).
 \label{11set3}
 \end{equation}

With these preliminaries we are prepared to prove the equivalence of dynamics contained in the following

Proposition: The dynamics of Maxwell Electrodynamics in Minkowski space-time can be equivalently described as a non-linear Electrodynamics in a curved space-time endowed with another metric written in terms of the electromagnetic field.

Let us consider the action defined in the e-space:

\begin{equation}
\hat{S} = \int \sqrt{- \hat{e}} \, \hat{L} \, d^{4}x
\label{3}
\end{equation}

To write this action in terms of the Minkowski metric we must insert in it the functions of the determinant of the e-metric and the Lagrangian $ \hat{L}$ in terms of the quantities $F$ and $ G$. The inverse functions, that is, $ F $ and $G$ in terms of $ \hat{F}$ and $ \hat{G}$ are very intricate algebraic expressions with which we will deal later on. For the time being we do not need to know their formulas explicitly. Indeed, consider that the Lagrangian $\hat{L}(\hat{F}, \hat{G}) $ is such that written in terms of the quantities $ F $  and $ G $ gives

$$ \hat{L}(\hat{F}, \hat{G}) =  \, U \, F.$$
Using the expression of the determinant the action (\ref{3}) goes into Maxwell dynamics
$$ \hat{S} = \int \sqrt{- \eta} \, F \, d^{4}x. $$

\vspace{0.50cm}
\textbf{An explicit example}

\vspace{0.50cm}
Consider the Lagrangians $\hat{L}_{1}$ and $\hat{L}_{2}:$

\begin{equation}
\hat{L}_{1} = \Omega^{- 2}
\label{31agosto1}
\end{equation}
and
\begin{equation}
\hat{L}_{2} = \Omega \, \hat{G}^{2}
\label{31agosto2}
\end{equation}

where the scalar $ \Omega$ is defined by

\begin{equation}
\Omega \, \equiv \, \frac{\sqrt{-e}}{\sqrt{-\eta}}
\label{4out1}
\end{equation}

Let us set for the dynamics the combined action

\begin{equation}
S = - \, \frac{\beta^{2}}{2} \left(\, S_{1} + \frac{1}{16 \, \beta^{4}} \, S_{2}\right) = - \, \frac{\beta^{2}}{2} \, \int \, \sqrt{-\hat{e}} \,\left( \hat{L}_{1} + \frac{1}{16 \, \beta^{4}} \, \hat{L}_{2} \right)
\label{31agosto3}
\end{equation}

This dynamics, when described in the Minkowski scenario is nothing but linear Maxwell electrodynamics. The proof is straightforward. Indeed, this action can be re-formulated as

$$ \delta \, S_{1} = - \, \frac{\beta^{2}}{2}  \,\delta \int \frac{\sqrt{-\eta}}{U} \, U^{2} \, d^{4}x $$

$$ \delta \, S_{2} = - \, \frac{\beta^{2}}{2}  \, \delta \int \frac{\sqrt{-\eta}}{U} \, U^{- 1} \, \hat{G}^{2}  \, d^{4}x $$

or

$$ \delta \, S_{2} = - \, \frac{\beta^{2}}{2}  \, \delta \int \frac{\sqrt{-\eta}}{U} \, U^{- 1} \, U^{2} \,G^{2} \, d^{4}x $$

Then

\begin{equation}
\delta S = - \, \frac{\beta^{2}}{2}  \, \delta \,  \int \, \sqrt{-\eta} \, \left( 1  + \frac{F}{2 \, \beta^{2}} \right) \,  d^{4}x
\label{31agosto4}
\end{equation}

which, apart from a cosmological term, reproduces Maxwell\rq s electrodynamics.

In the other way round we can use the relations
\begin{equation}
 F^{\mu\nu} = \frac{1}{\Sigma}  \, \hat{F}^{\mu\nu} - \frac{\Pi \, \Omega}{\Sigma} \, ^{*}\hat{F}^{\mu\nu}
\label{16set1}
\end{equation}
\begin{equation}
 ^{*}F^{\mu\nu} = \frac{1}{\Omega}  \, ^{*}\hat{F}^{\mu\nu}.
 \label{16set3}
 \end{equation}
where in the expressions for $ \Sigma $ and $ \Pi$ given by eq. (\ref{11set1}, \ref{11set3}) $F $ and $ G $ must be transformed in terms of $\hat{F},\hat{G}.$ The invariants $ F $ and $ G $  take the form
$$F =  \, 2 \, \beta^{2} \, \left( 1 - \frac{1}{\Omega} - \frac{1}{16 \, \beta^{4}} \, \hat{G}^{2} \, \Omega^{2} \right) $$
and
$$ G = \hat{G} \, \Omega. $$
Then we recognize immediately that the Lagrangian
$$
L_{M} =  2 \, \beta^{2} \, \left( \frac{1}{\Omega} - \frac{1}{\Omega^{2}} - \frac{1}{16 \, \beta^{4}} \, \hat{G}^{2} \, \Omega \right) $$
is nothing but Maxwell dynamics written in the e-metric.

\vspace{0.50cm}
\textbf{Topological invariant}

\vspace{0.50cm}
In the e-metric the topological invariant has the same expression as in the Minkowski metric. Indeed, set

 $$ I = \int \, \sqrt{-e} \, \hat{G} \, d^{4}x.$$
 Using the bridge relation we re-write this action as the topological invariant in Minkowski space-time:

 $$ I = \int \, \sqrt{-\eta} \, G \, d^{4}x. $$

\vspace{0.50cm}
\textbf{Inverse formula}

\vspace{0.50cm}

The inverse expressions of formulas (\ref{16set1}) and (\ref{16set3}) that relates $ F$ and $G$ to its correspondent in the e-metric are very involved. However we can obtain valuable information if we restrict our calculations up to order $O(1/\beta^{2}).$ We find

\begin{equation}
 F^{\mu\nu} \approx \hat{F}^{\mu\nu} - \frac{1}{2 \, \beta^{2}} \, \left[ \hat{F} \, \hat{F}^{\mu\nu} - \frac{\hat{G}}{2} \, ^{*}\hat{F}^{\mu\nu} \right]
\label{16set5}
\end{equation}
Thus

$$ F \approx \hat{F} - \frac{1}{\beta^{2}} \, ( \hat{F}^{2} + \frac{1}{2} \, \hat{G}^{2} )$$

$$ G \approx \hat{G} - \frac{1}{2 \, \beta^{2}} \, ( \hat{F} \, \hat{G}).$$

It then follows that Maxwell\rq s dynamics can be described in two equivalent ways:
\begin{itemize}
\item{Standard formulation in the Minkowski background:

$$\left( \sqrt{- \eta} \, F^{\mu\nu}\right)_{, \, \nu}= 0.$$}

\item{Alternative formulation in the e-metric:

$$\left( \sqrt{- \hat{e}} \,\left[ \frac{1}{\Omega \Sigma}\, \hat{F}^{\mu\nu} - \frac{\Pi}{\Sigma} \, ^{*}\hat{F}^{\mu\nu}\right] \right)_{, \, \nu}= 0$$}
\end{itemize}
where in the expressions of $ \Sigma $ and $ \Pi$ given by eq. (\ref{11set1}, \ref{11set3}) $F $ and $ G $ must be transformed in terms of $\hat{F},\hat{G}.$
\vspace{1.00cm}
Using the approximation up to order $ 1/\beta^{2}$ we obtain

$$ \left( \sqrt{- \hat{e}} \,\left[ \hat{F}^{\mu\nu} - \frac{1}{2 \, \beta^{2}} \, (\hat{F}\,\hat{F}^{\mu\nu}  - \hat{G} \, ^{*}\hat{F}^{\mu\nu}) \right] \right)_{, \, \nu} \approx 0 $$

\vspace{0.50cm}
\textbf{The source}

\vspace{0.50cm}

Let us now consider the case in which a current is introduced changing the dynamics into the form

\begin{equation}
\frac{1}{\sqrt{- \eta}} \, \left( \sqrt{- \eta} \, F^{\mu\nu}\right)_{, \, \nu}= J^{\mu}.
\label{16set9}
\end{equation}

which can be written in terms of the e-metric in the approximation we are considering as

\begin{equation}
  \frac{1}{\sqrt{- \hat{e}}} \, ( 1 - \frac{\hat{F}}{2 \, \beta^{2}}) \, \left[ \sqrt{- \hat{e}} \,\left( \hat{F}^{\mu\nu} - \frac{1}{2 \, \beta^{2}} \, (\hat{F}\,\hat{F}^{\mu\nu}  - \hat{G} \, ^{*}\hat{F}^{\mu\nu}) \right) \right]_{, \, \nu} \approx J^{\mu}.
  \label{16set11}
  \end{equation}
or, using the covariant derivative ($ \, ; \,$) it becomes

\begin{equation}
  \left( \hat{F}^{\mu\nu} - \frac{1}{2 \, \beta^{2}} \, (\hat{F}\,\hat{F}^{\mu\nu}  - \hat{G} \, ^{*}\hat{F}^{\mu\nu}) \right)_{ ; \, \nu} \approx \hat{J}^{\mu}
  \label{16set13}
  \end{equation}
where
$$  \hat{J}^{\mu} =  ( 1 + \frac{\hat{F}}{2 \, \beta^{2}} \,) \, J^{\mu}.$$

The conservation of the current can thus be expressed either in the Minkowski background
$$ J^{\mu}_{, \, \mu}=0 $$
or equivalently in the e-metric as

$$  \hat{J}^{\mu}_{; \, \mu}=0 $$
using the covariant derivative of the e-metric.

\subsection{Synthetic comments}
Let us pause for a while and consider what we have achieved. We have shown that there are two equivalent formulations of Maxwell Electrodynamics, to wit:
\begin{itemize}
\item{Maxwell linear action in a given frozen metric, say, Minkowski geometry $ \eta_{\mu\nu};$ }
\item{A non-linear action written in terms of an electromagnetic metric described in terms of the electromagnetic tensor.}
\end{itemize}
The description of Electrodynamics in either one of these approaches is just a matter of taste. We note that this equivalence is valid not only in the case of linear Maxwell theory but is true for any non-linear Electrodynamics that preserves gauge invariance. Besides, the background does not need to be restricted to Minkowski space-time but may be any metric $ g_{\mu\nu}.$

There is one consequence of this equivalence in the realm of the analysis of the test particles that we would like to explore here. The world of elementary particles shows that there are particles that do not interact with  the electromagnetic field. Another way to state this fact is to say that these particles do not have electric charge. This is an obvious statement in the standard gauge framework. However, in the e-metric formulation such identification is no more true. We can imagine, at least as a working hypothesis that it may be possible to generate a direct interaction of a neutral particle, say a neutrino, and the electromagnetic field that mimics the behavior of gravitational interaction. Indeed, we have learned from general relativity that the existence of a metric changes the properties of measuring distances and times in an universal way. Otherwise it is just a convenient tool to describe certain particular properties, as is the case of Gordon\rq s analysis of propagation of light in moving dielectrics or in Unruh\rq s sound propagation \cite{novellovisservolovik}.

Thus it seems worth to investigate the consequences of the universality hypothesis of the e-metric, representing the modification on all kind of matter imbedded in an electromagnetic field.

In this vein charged particles acquire two channels of interaction: the standard one (that needs a charge to establish the contact of the body with the electromagnetic field) and another one mediated by a modification of the geometry, a process that is made theoretically possible only after the introduction of the e-metric associated to the electromagnetic field. Assuming universality of the geometric structure implies that uncharged bodies do interact with the electromagnetic field trough the e-metric channel.

\section{Test particles in a given electromagnetic field: the case of fermions}

The possibility to describe electrodynamics in terms of a modification of the metric of space-time opens a new scenario to investigate the interaction of particles of any kind to the electromagnetic field. Let us explore this possibility for a fermion. In Minkowski space-time a free fermion is described by Dirac\rq s equation:

$$ i \, \gamma^{\mu} \, \partial_{\mu} \, \Psi - m \, \Psi =0,$$
where we take $\hbar = c = 1.$
In the presence of an electromagnetic field, the gauge principle states that the interaction of a fermion endowed with charge $ e $ is provided by the substitution

$$ \partial_{\mu} \rightarrow \partial_{\mu} - i\, e \, A_{\mu}.$$

This form of coupling requires the existence of a charge $ e.$ Equivalently, only charged particles interact with the electromagnetic field. However, the possibility to treat the  electromagnetic interaction in terms of a modification of the metric opens another theoretical treatment that is worth of examination.

First of all we recognize that the existence of a dimensionless metric allows the possibility to couple any particle to the electromagnetic field through the minimal coupling principle in a similar way as it is done in the gravitational interaction in the framework of general relativity. It is understood that a charged body have two channels of interaction: the standard gauge principle as stated above and the geometrical one through the interaction with the e-metric. Let us concentrate here only in the second mode. The starting point concerns the definition of an internal connection $ \hat{\Gamma}_{\mu} $ according to Fock and Ivanenko. Thus the minimal coupling of an uncharged fermion to an electromagnetic field takes the form
\begin{equation}
 i \, \hat{\gamma}^{\mu} \, \hat{\nabla}_{\mu}\, \Psi - m \,\Psi = 0.
 \label{7set3}
 \end{equation}
where the internal covariant derivative is given by

$$ \hat{\nabla}_{\mu} \,\Psi =\partial_{\mu} \, \Psi - \hat{\Gamma}_{\mu} \, \Psi$$

through the Fock-Ivanenko coefficient defined by \cite{brillwheeler}

\begin{equation}
 \hat{\Gamma}_{\mu} = - \, \frac{1}{8} \, \left( \hat{\gamma}^{\lambda} \,\hat{\gamma}_{\lambda \, , \mu}  -  \hat{\gamma}_{\lambda \, , \mu} \,\hat{\gamma}^{\lambda} -
\hat{\Gamma}^{\varrho}{}_{\mu\alpha}\, (\hat{\gamma}^{\alpha} \, \hat{\gamma}_{\varrho} - \hat{\gamma}_{\varrho} \,\hat{\gamma}^{\alpha}) \right)
\label{7set7}
\end{equation}
From the form of the metric
$$ \hat{e}^{\mu\nu} = \eta^{\mu\nu} - \frac{1}{\beta^{2}} \, \Phi^{\mu\nu}$$
where
$$ \Phi^{\mu\nu} = F^{\mu\alpha} \, F_{\alpha}{}^{\nu}.$$
Note that the gamma matrices satisfy the relation
$$ \hat{\gamma}^{\mu} \, \hat{\gamma}^{\nu} + \hat{\gamma}^{\nu} \, \hat{\gamma}^{\mu} = 2 \, \hat{e}^{\mu\nu} \, \mathbb{I}$$
where $ \mathbb{I} $ is the identity of the Clifford algebra. We can then write $ \hat{\gamma}^{\mu} $ in terms of the constant matrices of the Minkowski background $ \gamma^{\mu}:$

\begin{equation}
\hat{\gamma}^{\mu} = P^{\mu}{}_{\alpha} \, \gamma^{\alpha}
\label{7set9}
\end{equation}
where
$$  P^{\mu}{}_{\alpha} = \delta^{\mu}{}_{\alpha} - \frac{1}{\beta} \, F^{\mu}{}_{\alpha}.$$

The inverse covariant expression $ \hat{\gamma}_{\mu} = \hat{e}_{\mu\nu} \, \hat{\gamma}^{\nu}$ is given by

$$ \hat{\gamma}_{\mu} = \gamma^{\alpha} \, \left( A \, \eta_{\mu\alpha} - \frac{1}{\beta} \, ( A - \frac{F}{2 \, \beta^{2} \, U} ) F_{\mu\alpha} +
\frac{G}{4 \, \beta^{3} \, U}  \,  F^{*}_{\mu\alpha} + \frac{1}{\beta^{2} \, U} \, \Phi_{\mu\alpha}\right), $$
where $ A$ was defined in the inverse expression of the metric eq.(\ref{4out3}). To simplify notation we re-write it as

$$ \hat{\gamma}_{\mu} =   \mathbb{M}_{\mu\alpha} \, \gamma^{\alpha}.$$
Note that $ \mathbb{M}_{\mu\alpha}$ has no definite symmetry. Using this into the expression of the spinor connection we obtain
\begin{equation}
- \, 8 \, \hat{\Gamma}_{\mu} = P^{\lambda}{}_{\alpha} \left( \, \mathbb{M}_{\lambda\beta \, , \mu} - \hat{\Gamma}^{\varrho}_{\mu\lambda} \, \mathbb{M}_{\varrho\beta}\right) \, \Sigma^{\alpha\beta}
\label{8set5}
\end{equation}
where $ \Sigma^{\alpha\beta} = \gamma^{\alpha} \,\gamma^{\beta} - \gamma^{\beta} \, \gamma^{\alpha}.$ Now, we have
$$ \hat{\gamma}^{\mu} \, \hat{\Gamma}_{\mu} = - \frac{1}{8} \, \left(\gamma^{\mu} - \frac{1}{\beta} \, F^{\mu}{}_{\varepsilon} \,\gamma^{\varepsilon} \right) \, Y_{\alpha\beta\mu} \, \Sigma^{\alpha\beta} $$
where
$$ Y_{\alpha\beta\mu} = P^{\lambda}{}_{\alpha} \, \left( \mathbb{M}_{\lambda\beta ,\mu} -\hat{\Gamma}^{\varrho}_{\mu\lambda} \, \mathbb{M}_{\varrho\beta}\right) $$

Using the identity

\begin{equation}
 \gamma_{\mu} \, \Sigma_{\varrho\nu} = 2 \, \eta_{\mu\varrho} \, \gamma_{\nu}  - 2 \, \eta_{\mu\nu} \, \gamma_{\varrho} +2 \, i \, \varepsilon_{\mu\varrho\nu\sigma} \, \gamma^{5} \, \gamma^{\sigma}
 \label{19set1}
 \end{equation}

the equation of the uncharged fermion written in the Minkowski metric takes the form

\begin{equation}
i \, \gamma^{\mu} \, \partial_{\mu} \Psi + \frac{i}{\beta} \,F^{\mu\nu} \, \gamma_{\mu} \, \partial_{\nu} \Psi + i \,C_{\mu} \, \gamma^{\mu} \, \Psi - D_{\mu}  \gamma^{5} \, \gamma^{\mu} \,  \, \Psi - m \, \Psi = 0
\label{10set1}
\end{equation}
 where

 $$ C_{\mu} = \frac{1}{4} \,  \, \left( Y_{\alpha\mu}{}^{\alpha} - Y_{\mu\alpha}{}^{\alpha} + \frac{1}{\beta} Y_{\alpha\mu\beta} \, F^{\alpha\beta} - \frac{1}{\beta} Y_{\mu\varepsilon\lambda} \, F^{\varepsilon\lambda} \right) $$
 $$ D_{\mu} =  \frac{1}{4} \, \left( Y_{\alpha\beta\lambda} \, \varepsilon^{\alpha\beta\lambda}{}_{\mu} - \frac{1}{\beta} \, Y_{\alpha\beta\sigma} \, F^{\sigma}{}_{\varepsilon} \,\varepsilon^{\varepsilon\alpha\beta}{}_{\mu} \,\right). $$

We find three forms of coupling between an uncharged fermion and the electromagnetic field, that we can synthesize in the following way:

$$C_{\mu} \, \bar{\Psi} \, \gamma^{\mu} \, \Psi;$$
$$F^{\mu\nu} \, \bar{\Psi} \, \gamma_{\mu} \,\partial_{\nu} \Psi; $$
$$D_{\mu} \, \bar{\Psi} \, \gamma^{\mu} \, \gamma^{5} \,\Psi.$$
It is immediate to recognize that without any limit for the electromagnetic field, that is, in the present framework, to take the limit $ \beta \, \rightarrow \, \infty $ all these three forms of coupling disappears. This is the case of standard Maxwell theory: no limit for the energy of the field and consequently no electromagnetic interaction for the uncharged particle.

\section{Back-reaction: the influence of uncharged fermions on the dynamics of the electromagnetic field}

In the previous section we described the influence of the electromagnetic field on uncharged fermions. Let us now consider the back-reaction that affects
the dynamics of the electromagnetic field. To simplify our presentation (and due to the fact that the value of $ \beta$ is very high) we will limit our analysis to second order $ O(1/ \beta^{2}.)$

We set the total action  $ S = S_{EM} + S_{\Psi}. $
In the Minkowski background the free Lagrangian for the electromagnetic field is given by
$$ L_{EM} = - \, \frac{1}{4} \, F.$$
Assuming the minimal coupling of the fermion with the e-metric provides the following action for the fermion
\begin{equation}
S_{\Psi} = \int \, \sqrt{- \hat{e}} \, \left( \frac{i}{2} \, \bar{\Psi} \, \hat{\gamma}^{\mu} \, \hat{\nabla}_{\mu} \Psi - \frac{i}{2} \,\hat{\nabla}_{\mu} \bar{\Psi} \, \hat{\gamma}^{\mu} \,  \Psi - m \, \bar{\Psi} \, \Psi \right)
\label{17set1}
\end{equation}

In order to re-write this interaction in the Minkowski background we list the approximations of the relations that we need to obtain the covariant derivative and the determinant of the e-metric. They are:

$$ \sqrt{- \, \hat{e}} \approx \sqrt{- \eta} \, ( 1 - \frac{F}{2 \, \beta^{2}}) \approx \sqrt{- \eta} \, ( 1 - \frac{\hat{F}}{2 \, \beta^{2}})$$

$$ G \approx \hat{G} - \frac{\hat{F} \, \hat{G}}{2 \, \beta^{2}}  $$

$$ A = \frac{2 + F/\beta^{2}}{2 \, U} \approx 1 $$

$$ \mathbb{M}_{\mu\nu} \approx \eta_{\mu\nu} - \frac{1}{\beta} \, F_{\mu\nu}  + \frac{1}{\beta^{2}} \, \Phi_{\mu\nu}$$

$$ \hat{e}_{\mu\nu} \approx \eta_{\mu\nu}+ \frac{1}{\beta^{2}} \, \Phi_{\mu\nu} $$

$$ \hat{\gamma}_{\mu}  \approx \gamma_{\mu} - \frac{1}{\beta} \, F_{\mu}{}^{\alpha} \, \gamma_{\alpha} + \frac{1}{\beta^{2}} \, \Phi_{\mu}{}^{\alpha} \, \gamma_{\alpha} $$

$$ 8 \, \hat{\Gamma}_{\mu} \approx \left( \frac{1}{\beta} \, F_{\alpha\beta \, , \mu} - \frac{1}{\beta^{2}} \, ( F_{\lambda\alpha} \, F^{\lambda}{}_{\beta \, , \mu} + \Phi_{\alpha\mu \, , \beta} \,) \right) \,  \Sigma^{\alpha\beta}. $$

From these expressions we obtain

$$ \hat{\Gamma}^{\varrho}{}_{\mu\alpha} \, (\hat{\gamma}^{\alpha} \, \hat{\gamma}_{\varrho} -\hat{\gamma}_{\varrho} \, \hat{\gamma}^{\alpha} ) \approx \frac{1}{\beta^{2}} \,\Phi_{\alpha\mu \, , \, \beta} \, \Sigma^{\alpha\beta} $$

Then eq. (\ref{17set1}) becomes

\begin{equation}
S_{\Psi} = \int \, \sqrt{- \eta} \, \left( \mathbb{L}_{1} + \, \mathbb{L}_{2} + \mathbb{L}_{3}  \right)
\label{18set1}
\end{equation}

where
$$
\mathbb{L}_{1} \approx  ( 1 - \frac{F}{2 \, \beta^{2}} ) \left( \frac{i}{2} \, \bar{\Psi} \, \gamma^{\mu} \, \partial_{\mu} \Psi - \frac{i}{2} \, \partial_{\mu} \bar{\Psi} \, \gamma^{\mu} \,  \Psi - m \, \bar{\Psi} \, \Psi \right) $$

$$ \mathbb{L}_{2} \approx  \, \frac{i}{2\, \beta}\, [ \bar{\Psi} \, \gamma_{\alpha} \, {\partial}_{\beta} \Psi -  \, \partial_{\beta} \bar{\Psi} \, \gamma_{\alpha} \,  \Psi] \, F^{\alpha\beta} $$

$$ \mathbb{L}_{3} \approx i \, \left( - \, Z_{\mu\alpha\beta} + \,  P_{\mu\alpha\beta} \right) \, \left( \bar{\Psi} \, \gamma^{\mu} \, \Sigma^{\alpha\beta} \,  \Psi  + \bar{\Psi} \,  \Sigma^{\alpha\beta} \, \gamma^{\mu} \, \Psi\right)$$

and in the approximation up to order $O(1/\beta^{2}) $ we have

$$ Z_{\mu\alpha\beta} \approx \frac{1}{16 \, \beta} \, F_{\alpha\beta \, , \mu} $$

 $$ P_{\mu\alpha\beta}  \approx  \frac{1}{16 \, \beta^{2}} \left( F_{\lambda\alpha} \, F^{\lambda}{}_{\beta \, , \mu} + \Phi_{\alpha\mu \, , \beta} + F^{\lambda}{}_{\mu} \, F_{\alpha\beta \, , \lambda} \right) $$

which provides the modification on the dynamics of the electromagnetic field due to the uncharged particles.

\section{Final comments}

The description of Maxwell electrodynamics in terms of a special modification of the metric of space-time opens a new possible interaction of particles with the electromagnetic field besides the standard gauge through the form $ \partial_{\mu} - i \, e \, A_{\mu}, $  a procedure that can be used only to charged particle. The method of the electromagnetic metric to describe Maxwell electrodynamics allows another possibility concerning both, charged and uncharged particles. In this paper we suggest that the electromagnetic metric could be used as a new channel of interaction for all particles with the electromagnetic field. This universality, which seems weird at first glance, could be conciliated with observations once the new effects appear only for very high values of the electromagnetic field. We have shown how this hypothesis can be analyzed. The formal existence of the electromagnetic metric requires a bound for the maximal value of the field which we named $ \beta.$ The present model is not able to precise the value of $ \beta$ \cite{schw}. If there is no superior limit for the electromagnetic field, setting $ \beta$ to $\infty$ corresponds to the standard formulation of Electrodynamics.

\section*{Acknowledgements}
 This work was partially supported by {\em Conselho Nacional de Desenvolvimento
Cient\'{\i}fico e Tecnol\'ogico} (CNPq) and \emph{Coordena\c{c}\~ao do Aperfei\c{c}oamento do Pessoal do Ensino Superior (CAPES)}. We would like to thank Ugo Moschella for a critical reading of a previous version of this paper.

\section{Appendix: Charged bodies in an electromagnetic field}
Let us consider now what are the modifications of the electromagnetic force on a test particle endowed with velocity $ v_{\mu}$ due to the presence of the associated e-metric.
We set
$$ \hat{v}_{\mu} =  v_{\mu}.$$
The contra-variant form is given by
$$  \hat{v}^{\mu} =  \hat{e}^{\mu\nu} \, \hat{v}_{\nu} = ( 1 - \frac{E^{2}}{\beta^{2}} ) \, v^{\mu} - \frac{2}{\beta^{2}} \, q^{\mu}$$
where $ q^{\mu} = \frac{1}{2} \, \eta^{\mu\nu\alpha\beta} \, E_{\nu} \, v_{\alpha} \, H_{\beta}$ is the heat flux (Poynting vector). We remind that for any non-hat tensorial quantity all properties concerning the relationship between covariant and contra-variant indices are performed by Minkowski metric $ \eta_{\mu\nu}.$

In the standard formulation of Maxwell\rq s theory in Minkowski space-time an uncharged particle does not interact with the electromagnetic field. The gauge principle that guide such interaction needs a dimensional quantity, the charge, to implement this interaction. Thus, there is no room for coupling directly an uncharged particle to the electromagnetic field. However, from the knowledge of the e-metric a new possibility appears. In order to analyze the motion of the particle in the e-metric  we proceed starting from first principles and using directly a formal expression to analyze the acceleration suffered by a charged or an uncharged particle. We make appeal to the intimate relationship between the field and the metric. Indeed the acceleration is given by
\begin{equation}
\hat{a}_{\mu} = \hat{v}_{\mu \, ; \, \nu} \, \hat{v}^{\nu} = \left( \hat{v}_{\mu \, , \, \nu} - \hat{\Gamma}^{\alpha}_{\mu\nu} \, \hat{v}_{\alpha} \right) \, \hat{v}^{\nu}.
\label{3set1}
\end{equation}
We can then proceed and develop the connection to obtain
$$ \hat{\Gamma}^{\alpha}_{\mu\nu} \, \hat{v}_{\alpha} \, \hat{v}^{\nu} = \frac{1}{2} \, \hat{e}_{\lambda\nu \, , \mu} \, \hat{v}^{\lambda} \, \hat{v}^{\nu} $$

Let us deal with the simplest case where we can set $ q_{\mu} = 0 $ and
$$ v_{\mu \, , \nu} = a_{\mu} \, v_{\nu} $$

Using the inverse metric $ \hat{e}_{\mu\nu}$  we have

\begin{equation}
\hat{e}_{\lambda\nu \, , \mu} \, \hat{v}^{\lambda} \, \hat{v}^{\nu} = ( 1 - \frac{E^{2}}{\beta^{2}})^{2} \, \left( X_{, \, \mu} - 2 \, \hat{e}_{\lambda\nu} \, a^{\lambda} \, v^{\nu} \, v_{\mu} \right)
\label{9set1}
\end{equation}
where
$$ X \equiv \frac{1}{U} \, \left( 1 + \frac{H^{2}}{\beta^{2}}\right).$$

Then the acceleration in the e-metric is given by:

\begin{equation}
\hat{a}_{\mu} = ( 1 - \frac{E^{2}}{\beta^{2}} ) \,  a_{\mu}  -  \frac{1}{2} \, (1 - \frac{E^{2}}{\beta^{2}})^{2} \,  X_{ , \mu}.
\label{6set4}
\end{equation}
where $ a_{\mu} = q \, F_{\mu\nu} \, v^{\nu}$  is the acceleration in the limit $ \beta \rightarrow \infty$ and $ q $ is the charge.
Then we can re-write the acceleration up to order $ O(1 / \beta^{2})$ as

\begin{equation}
\hat{a}_{\mu} = \frac{q}{m} \, \hat{F}_{\mu\nu} \,  \hat{v}^{\nu}  -  \frac{1}{2 \,\beta^{2}} \, (E^{2} + H^{2})_{ , \, \mu}
\label{6set5}
\end{equation}
where we are using $ c = 1.$ This is the form of action of the electromagnetic field on a test particle. Note that correction on the Lorentz force appears only if the field is high enough, that is, if we can not neglect terms of order $  E^{2}/\beta^{2}.$ For uncharged bodies only the second term of (\ref{6set5}) survives. If there is no limit for the values of the field and we can take $ \beta \rightarrow \infty $ this formula reduces to the Lorentz force.

 \end{document}